\documentclass[twocolumn,prl,aps]{revtex4}

\begin{document}
\title{Local Quantum Pure-state Identification without Classical Knowledge}
\author{Y. Ishida, T. Hashimoto, M. Horibe, and A. Hayashi}
\affiliation{Department of Applied Physics\\
           University of Fukui, Fukui 910-8507, Japan}

\begin{abstract}
Suppose we want to distinguish two quantum pure states. We consider the case in which 
no classical knowledge on the two states is given and only a pair of samples of 
the two states is available. This problem is called quantum pure-state identification 
problem. Our task is to optimize the mean identification success probability, 
which is averaged over an independent unitary invariant distribution of the two 
reference states. In this paper, the two states are assumed bipartite states which are 
generally entangled. The question is whether the maximum mean identification success probability 
can be attained by means of an LOCC (Local Operations and Classical Communication) measurement 
scheme. We will show that this is possible by constructing a POVM which respects the 
conditions of LOCC.
\end{abstract}

\maketitle

\newcommand{\ket}[1]{|\,#1\,\rangle}
\newcommand{\bra}[1]{\langle\,#1\,|}
\newcommand{\braket}[2]{\langle\,#1\,|\,#2\,\rangle}
\newcommand{\bold}[1]{\mbox{\boldmath $#1$}}
\newcommand{\sbold}[1]{\mbox{\boldmath ${\scriptstyle #1}$}}
\newcommand{\tr}[1]{{\rm tr}\!\left[#1\right]}
\newcommand{\trm}{{\rm tr}}
\newcommand{\BC}{{\bold{C}}}
\newcommand{\CS}{{\cal S}}
\newcommand{\CA}{{\cal A}}
\newcommand{\CM}{{\cal M}}

\section{Introduction}
It is an extremely nontrivial problem to distinguish different states of a 
quantum system by measurement \cite{Helstrom76,Ivanovic87,Dieks88,Peres88}. 
First of all, this is because of statistical 
nature of quantum measurement, which destroys the state of the system 
to be measured and does not allow one to clone an unknown quantum state \cite{Wootters82}. 
Another relevant issue is nonlocality of quantum mechanics. When the system to be measured is a 
composite, we can generally obtain more information of 
the system by the global measurement on the whole system than by a combination of local 
measurements on its subsystems \cite{Bennett99, Koashi07}.

Let us focus on the problem of distinguishing two pure states of a composite 
system which is shared by two parties. Is is a fundamental 
question of quantum information theory whether the optimal distinguishment 
can be performed by means of local operations and classical 
communication (LOCC) scheme of the two parties.

Walgate {\it et al}. \cite{Walgate00} showed that any two mutually orthogonal pure states can be 
perfectly distinguished by LOCC. This is rather surprising 
since their result holds regardless of entanglement of the states, 
which is a typical source of nonlocality in quantum information.  
It has also been shown that any two generally nonorthogonal pure states 
can be optimally discriminated by LOCC: the optimal 
success probability of discrimination by the global measurement can be 
attained by an LOCC protocol. This was shown for the two types of  
discrimination problems: the inconclusive discrimination problem \cite{Virmani01} 
where error is allowed and the conclusive (unambiguous) discrimination 
problem \cite{Ji05,Chen01,Chen02} where no error is allowed but an inconclusive 
guess can be made.  These results can be interpreted that there is no 
nonlocality in the discrimination of two pure states.

We can consider a different setting for discrimination problem of two pure 
states. In the usual setting, it is assumed that perfect classical knowledge 
of the two states $\rho_1$ and $\rho_2$ is given to the two parties. 
The measurement scheme for the optimal discrimination naturally depends on 
the classical knowledge of the states.  
Instead, let us assume that no classical knowledge of the states 
$\rho_1$ and $\rho_2$ are given, but a certain number ($N$) of their copies 
are available as reference states. 
One's task is correctly identify a given input state $\rho$ with one of 
the reference states $\rho_1$ and $\rho_2$ by means of a measurement on the 
whole state $\rho \otimes \rho_1^{\otimes N} \otimes \rho_2^{\otimes N}$.
When the number of copies $N$ is infinite, the problem is reduced to quantum 
state discrimination. This is because we can always obtain complete classical 
knowledge of a quantum state if we have infinitely many copies of 
the state. 
We call this problem "quantum state identification".  The optimal 
success probability has been determined for the inconclusive \cite{Hayashi05} and 
conclusive (unambiguous) \cite{Bergou05, Hayashi06} identification problems. 

In this paper, we investigate the inconclusive pure-state identification 
problem of $N=1$ where the two reference pure states $\rho_1$ and $\rho_2$ are bipartite. 
The input state $\rho$ given to Alice and Bob is guaranteed to be one of the 
reference states $\rho_1$ and $\rho_2$ with an {\it a priori} probability 
$\eta_1$ and $\eta_2$. The two reference states are assumed to be independently 
distributed on the pure state space in a unitary invariant way. 
Each reference state generated this way is generally entangled. 
We will demonstrate that Alice and Bob can identify the input state by means of an LOCC 
protocol with the success probability given by the optimal global identification 
scheme.   

\section{Pure-state identification problem without LOCC conditions}
In this section, we will precisely formulate the pure-state identification 
problem and derive the maximum mean success probability without the LOCC conditions 
for the case of $N=1$ and an arbitrary {\it a priori} occurrence probability of 
the reference states. 
In the case of single-qubit system, the problem has been solved by Bergou {\it et al}. \cite{Bergou06}. 
For the case of general $N$ but with equal {\it a priori} 
occurrence probabilities, see Ref. \cite{Hayashi05}. 

We have three quantum systems numbered 0, 1, and 2, each on a $d$ dimensional 
space $\BC^{d}$.  The input pure state $\rho=\ket{\phi}\bra{\phi}$ is prepared in system 0 and 
the two reference pure states $\rho_1=\ket{\phi_1}\bra{\phi_1}$ and 
$\rho_2=\ket{\phi_2}\bra{\phi_2}$ in system 1 and 2, respectively. 
The space which an operator acts on is specified by the number in the parenthesis. 
For example, $\rho_1(1)$ is a density operator on system 1. 
The input state $\rho$ is promised to be one of the reference states $\rho_1$ and $\rho_2$ 
with an {\it a priori} probability $\{\eta_1, \eta_2\}$. 
The two reference states are independently chosen from the state space $\BC^{d}$ in  
a unitary invariant way. More precisely, the distribution is assumed uniform 
on the $2d-1$ dimensional unit hypersphere of $2d$ real variables 
$\{{\rm Re} c_i, {\rm Im} c_i\}_{i=0}^{d-1}$, 
where $c_i$ is expansion coefficients of the state 
in terms of an orthonormal base $\{ \ket{i} \}_{i=0}^{d-1}$. The distribution 
does not depend on a particular choice of the base.

Our task is correctly identify the input state with one of the reference states 
$\rho_\mu (\mu=1,2)$ by measuring the whole system $0 \otimes 1 \otimes 2$. 
We denote corresponding POVM elements by $E_\mu (\mu=1,2)$. The mean identification 
success probability is then given by 
\begin{eqnarray}
  p(d) = \sum_{\mu=1,2} \eta_\mu \Big<
           \tr{E_\mu \rho_\mu(0)\rho_1(1)\rho_2(2)} \Big>,
\end{eqnarray}
where the symbol $< \cdots  >$ represents the average over the reference states $\rho_1$ and 
$\rho_2$. Note that the POVM $E_\mu$ is independent of $\rho_1$ and $\rho_2$, since 
we have no classical knowledge on the reference states.  

The average over the reference states can be readily performed by using the formula \cite{Hayashi04}:
\begin{eqnarray}
  < \rho^{\otimes n} > = \frac{\CS_n}{d_n},
\end{eqnarray}
where $\CS_n$ is the projector on to the totally symmetric subspace of $(\BC^d)^{\otimes n}$ 
and $d_n$ is its dimension given by $d_n={}_{n+d-1}C_{d-1}$.
Using $E_2=1-E_1$, the mean success probability to be maximized is written as  
\begin{eqnarray}
  p(d) = \eta_2 + \frac{1}{d_1 d_2} \tr{ E_1( \eta_1\CS(01) - \eta_2\CS(02) ) }, \label{pd}
\end{eqnarray}
where $\CS(01)$ and $\CS(02)$ are the projector onto the totally symmetric subspace of 
space $0 \otimes 1$ and $0 \otimes 2$, respectively.  

The only restriction on the POVM element $E_1$ is $0 \le E_1 \le 1$. In order to maximize 
the mean success probability Eq.(\ref{pd}), we use the following result which holds for any Hermitian 
operator $\Delta$: 
\begin{eqnarray}
    && \max_{0 \le E \le 1} \tr{E\Delta} \nonumber \\
    && = \mbox{sum of all positive eigenvalues of $\Delta$},
\end{eqnarray}
where the maximum is attained when $E$ is the projector $P_+$ onto the subspace $V_+(\Delta)$ 
spanned by all eigenstates of $\Delta$ with a positive eigenvalue. 
Note that it does not matter whether the subspace $V_+(\Delta)$ includes eigenvectors with 
zero-eigenvalue.  
In our case, $\Delta$ is defined to be 
\begin{eqnarray}
   \Delta= \eta_1\CS(01) - \eta_2\CS(02). \label{delta}
\end{eqnarray}

Let us decompose the total space into three subspaces according to the symmetry with 
respect to system permutations \cite{Hamermesh62}. 
\begin{eqnarray}
V=\BC^d \otimes \BC^d \otimes \BC^d=V_\CS \oplus V_\CA \oplus V_\CM.
\end{eqnarray}
Here $V_\CS$ is the totally symmetric subspace of dimension $\dim V_\CS \equiv d_3=d(d+1)(d+2)/6$ 
and $V_\CA$ is the totally antisymmetric subspace of dimension $\dim V_\CM =d(d-1)(d-2)/6$. 
And the remaining subspace $V_\CM$ is the mixed symmetric subspace of dimension 
$\dim V_\CM =2d(d^2-1)/3$.  
The subspace $V_\CM$ contains the 2 dimensional irreducible representation of the symmetric group 
of order 3, $S_3$, with multiplicity $\dim V_\CM /2$. 
We will not exploit any representation theory of the symmetric group in the following arguments.
We denote projectors onto 
$V_\CS$, $V_\CA$, and $V_\CM$ by $\CS_3$, $\CA_3$ and $\CM_3$, respectively. 

It is clear that $\Delta=\eta_1-\eta_2$ in $V_\CS$ and $\Delta=0$ in $V_\CA$. 
To determine eigenvalues of $\Delta$ in $V_\CM$, it is convenient to introduce two 
operators $D$ and $A$ as  
\begin{eqnarray}
  &&  D \equiv \CS(01)-\CS(02) = \frac{1}{2} \left( T(01) - T(02) \right),  \label{D} \\
  &&  A \equiv \CS(01)+\CS(02)-1 = \frac{1}{2} \left(  T(01) + T(02) \right). \label{A}
\end{eqnarray}
Here, $T(01)$ is the operator which exchanges system 0 and 1 and $T(02)$ exchanges 
system 0 and 2. Calculating $D^2$, we find 
\begin{eqnarray}
   D^2 &=& \frac{1}{4} \left( 2-T(01)T(02)-T(02)T(01) \right) \\
       &=& \frac{3}{4} \left(1 - \CS_3 -\CA_3 \right) \\
       &=& \frac{3}{4} \CM_3,
\end{eqnarray}
which implies that eigenvalues of $D$ is $\pm \sqrt{3}/2$ in $V_\CM$ and 0 otherwise. 
It is also easy to show that 
\begin{eqnarray} 
  && DA+AD=0, \label{anti} \\
  && A^2 = 1-D^2. \label{asqure}
\end{eqnarray}
The anticommutability of Eq.(\ref{anti}) implies that if $\ket{+}$ is an eigenstate of $D$ 
with eigenvalue $\sqrt{3}/2$, then $A \ket{+}$ is also an eigenstate of $D$ with eigenvalue 
$-\sqrt{3}/2$. By Eq.(\ref{asqure}), we find that $\ket{-} \equiv 2A\ket{+}$ is correctly 
normalized. Note that the positive and negative eigenvalues of $D$ have the same 
multiplicity. 
Thus we can choose the orthonormal base $\{ \ket{+,k}, \ket{-,k} \}$ 
in $V_\CM$ such that 
\begin{eqnarray}
  && D\ket{+,k} = + \frac{\sqrt{3}}{2}\ket{+,k}, \\ 
  && D\ket{-,k} = - \frac{\sqrt{3}}{2}\ket{-,k}, \\ 
  && A\ket{+,k} = \frac{1}{2}\ket{-,k}, \\
  && A\ket{-,k} = \frac{1}{2}\ket{+,k},
\end{eqnarray}
where the index $k$ runs from 1 to $\dim V_\CM/2$.
In this base, $D$ and $A$ are block-diagonalized with respect to $k$ and 
each block has the following 2 by 2 matrix representation.
\begin{eqnarray}
  D = \left( \begin{array}{cc}
                 \frac{\sqrt{3}}{2}  &   0                  \\
                 0                   &  -\frac{\sqrt{3}}{2} \\
             \end{array}
      \right),\ \ 
  A = \left( \begin{array}{cc}
                 0           &  \frac{1}{2}  \\
                 \frac{1}{2} &  0            \\
             \end{array}
      \right).
\end{eqnarray}

In terms of $D$ and $A$, the operator $\Delta$ is written as 
\begin{eqnarray}
  \Delta = \frac{1}{2} \left( \eta_1-\eta_2 + D + (\eta_1-\eta_2)A \right). \label{deltaDA}
\end{eqnarray}
The operator $\Delta$ is also block-diagonalized with the same 2 by 2 matrix representation 
which can be readily diagonalized. Two eigenvalues of $\Delta$ are given by
\begin{eqnarray}
   \lambda_{\pm} = \frac{1}{2} \left( \eta_1-\eta_2 \pm \sqrt{1-\eta_1\eta_2} \right), 
\end{eqnarray}
and we find that  $\lambda_+ \ge 0$ and $\lambda_- \le 0$. 

Now we can calculate the maximum success probability. Let us assume $\eta_1 \ge \eta_2$ 
for the moment. The positive eigenvalues of $\Delta$ are $\eta_1 - \eta_2$ in $V_\CS$ 
with multiplicity $\dim V_\CS$ and $\lambda_+$ in $V_\CM$ with multiplicity 
$\dim V_\CM/2$. 
We thus obtain 
\begin{eqnarray}
  && p_{\max}(d) \nonumber \\
  && = \eta_2 + \frac{1}{d_1d_2} 
          \left( 
              (\eta_1 - \eta_2) \dim V_\CS + \lambda_+\frac{\dim V_\CM }{2} 
          \right) \nonumber \\
  && = \frac{1}{2} + \frac{d+2}{6d} (\eta_1-\eta_2) + \frac{d-1}{3d}\sqrt{1-\eta_1 \eta_2}.
\end{eqnarray}

If $\eta_1 \le \eta_2$, the only positive eigenvalue of $\Delta$ is $\lambda_+$ in $V_\CM$, 
hence we obtain  
\begin{eqnarray}
  && p_{\max}(d)  
        = \eta_2 + \frac{1}{d_1d_2}\lambda_+\frac{\dim V_\CM }{2} \nonumber \\
  && = \frac{1}{2} - \frac{d+2}{6d} (\eta_1-\eta_2) + \frac{d-1}{3d}\sqrt{1-\eta_1 \eta_2}.
\end{eqnarray}

These two cases can be combined to yield a symmetric form of 
the maximum success identification probability for 
general magnitude relation between $\eta_1$ and $\eta_2$. 
\begin{eqnarray}
  p_{\max} (d) = \frac{1}{2} + \frac{d+2}{6d} |\eta_1-\eta_2| + \frac{d-1}{3d}\sqrt{1-\eta_1 \eta_2}.
              \label{pmax}
\end{eqnarray}
The maximum is attained when the POVM element $E_1$ is given by $P_+$, the projector 
onto the subspace of positive eigenvalues of $\Delta$.  
The $p_{\max} (d)$ given by Eq.(\ref{pmax}) reproduces the result for the case $d=2$ obtained 
in Ref.\cite{Bergou06} and the one for arbitrary $d$ in Ref.\cite{Hayashi05}
when $\eta_1=\eta_2=1/2$. 

\section{Pure-state identification by LOCC}
Let us assume that each of the three systems 0, 1, and 2 , where the input state and the 
two reference states are prepared, consists of two subsystems. 
The state space of each system is represented by a tensor product 
$\BC^d=\BC^{d_a}\otimes \BC^{d_b}$, which is shared by Alice and Bob.  
Their task is to identify a given input bipartite state 
with one of the two bipartite reference states by means of local operations and classical 
communication (LOCC). As in the preceding section, the two reference states are chosen randomly 
from the pure state space $\BC^d$ in the unitary invariant way. 
Therefore, those bipartite states are generally entangled. 
The question is whether Alice and Bob can achieve the maximum mean identification success 
probability given by the global measurement scheme. In this section, we will show that 
this is possible by explicitly constructing an LOCC protocol which achieves it.  

The mean success probability is given by Eq.(\ref{pd}) in the preceding section. 
The optimal global POVM element $E_1$ is $P_+$, the projector onto 
the subspace of positive eigenvalues of $\Delta$ defined by Eq.(\ref{delta}).  
The projector $P_+$ does not apparently satisfy the conditions of LOCC, 
since the operator $\Delta$ is not of a separate form. 
However, it should be noticed that $\tr{E_1\Delta}$ remains the same if the support 
of $E_1$ contains states with zero-eigenvalue of $\Delta$. It is this freedom that 
we will exploit in order to construct a POVM element $E_1$ which satisfies the 
LOCC conditions.  

We begin with rewriting the operator $\Delta$ of Eq.(\ref{deltaDA}) in terms of 
local operators of Alice and Bob. 
Note that the exchange operator $T(01)$, for example, can be written as 
$T(01)=T^{(a)}(01) \otimes T^{(b)}(01)$, where $T^{(a)}(01)$ is the operator which exchanges Alice's 
part of system 0 and 1 and $T^{(b)}(01)$ is defined for Bob's part in the same way. Hereafter, 
we use the suffix $(a)$ or $(b)$ for an operator to indicate which space of Alice or Bob 
the operator acts on. Since we have 
\begin{eqnarray}
   D &=& D^{(a)}\otimes A^{(b)} + A^{(a)}\otimes D^{(b)},   \\
   A &=& D^{(a)}\otimes D^{(b)} + A^{(a)}\otimes A^{(b)},  
\end{eqnarray}
the operator $\Delta$ is expressed as
\begin{eqnarray}
   \Delta &=& \frac{1}{2} \Big(
               \eta_1-\eta_2 + D^{(a)}A^{(b)} + A^{(a)}D^{(b)}   \nonumber \\ 
          & & \ \ \ \ \ \ +(\eta_1-\eta_2) (D^{(a)}D^{(b)} + A^{(a)}A^{(b)}) \Big).
\end{eqnarray}

The task for Alice and Bob is to maximize $\tr{E_1^{{\rm L}}\Delta}$ with a POVM element  
$E_1^{{\rm L }}$ which satisfies LOCC conditions. 
We first construct a separable POVM $E_1^{{\rm L}}$ which 
attains the maximum value $\tr{P_+\Delta}$. This separable POVM $E_1^{{\rm L}}$ 
will then be shown to satisfy the LOCC conditions. 
Without loss of generality, we assume $\eta_1 \le \eta_2$ throughout this section, since the problem 
is symmetric with respect to $\rho_1$ and $\rho_2$. 

Suppose that Alice and Bob first determine the permutation symmetry of their systems by 
the projective measurement with projection operators 
$\{\CS_3^{(a)}, \CA_3^{(a)}, \CM_3^{(a)}\}$ and $\{\CS_3^{(b)}, \CA_3^{(b)}, \CM_3^{(b)}\}$, 
respectively.  
If one of them found that his or her system is totally symmetric or antisymmetric, 
it is easy for the other party to find the best strategy. 
For example, assume that Alice found her system to be totally symmetric.  
Knowing Alice's outcome, Bob performs a POVM measurement, which we denote by $x^{(b)}$. 
The contribution to $\tr{E_1^{{\rm L}}\Delta}$ is then given by
\begin{eqnarray}
  \tr{\CS_3^{(a)}\otimes x^{(b)} \Delta} = \dim (V_\CS^{(a)}) \trm_b[x^{(b)}\Delta^{(b)}], 
\end{eqnarray}
since $\CS_3^{(a)}D^{(a)}=0$ and $\CS_3^{(a)}A^{(a)}=\CS_3^{(a)}$. 
It is clear that the best strategy for Bob is to take the projector $P_+^{(b)}$ 
onto the positive-eigenvalue space of $\Delta^{(b)}$.  
Note that the positive-eigenvalue space of $\Delta^{(b)}$ is a subspace of 
$V_\CM^{(b)}$, since the eigenvalue of $\Delta^{(b)}$ in $V_\CS^{(b)}$ 
is $\eta_1 - \eta_2 (\le 0)$.  
In this case the contribution to $\tr{E_1^{{\rm L}} \Delta}$ is given by 
\begin{eqnarray}
   \tr{\CS_3^{(a)} \otimes P_+^{(b)} \Delta} = 
           \frac{\lambda_+}{2} \dim V_{\CS}^{(a)} \dim V_{\CM}^{(b)}.
\end{eqnarray}
If Alice's part is totally antisymmetric, the operator 
for Bob is given by 
\begin{eqnarray}
  && \trm_a[\CA_3^{(a)}\Delta] =  \dim (V_\CA^{(a)}) \Delta^{'(b)}, \nonumber \\ 
  && \Delta^{'(b)} \equiv \frac{1}{2}\left( \eta_1-\eta_2 - D^{(b)} -(\eta_1-\eta_2)A^{(b)} \right).
\end{eqnarray}
The operator $\Delta^{'(b)}$ differs from $\Delta^{(b)}$ only in the signs 
in front of $D^{(b)}$ and $A^{(b)}$. Its eigenvalues are 0 in $V_\CS^{(b)}$ 
and $\eta_1-\eta_2 (\le 0)$ in $V_\CA^{(b)}$. In $V_\CM^{(b)}$, the operator 
$\Delta^{'(b)}$ has eigenvalue $\lambda_-$ 
in the positive-eigenvalue subspace of $\Delta^{(b)}$ and $\lambda_+$ in the negative-eigenvalue 
subspace of $\Delta^{(b)}$. This implies Bob's best POVM element is $P_-^{(b)}$, the projector 
onto the $\Delta^{(b)}$'s negative-eigenvalue subspace in $V_\CM^{(b)}$.
The contribution to $\tr{E_1^{{\rm L}} \Delta}$ in this case is given by
\begin{eqnarray}
  \tr{\CA_3^{(a)} \otimes P_-^{(b)} \Delta} = 
           \frac{\lambda_+}{2} \dim V_{\CA}^{(a)} \dim V_{\CM}^{(b)}.
\end{eqnarray}
The same argument also holds when Bob's system is totally symmetric or antisymmetric.  
Therefore, when the total state does not belong to $V_\CM^{(a)}\otimes V_\CM^{(b)}$, 
the whole contribution to $\tr{E_1^{{\rm L}} \Delta}$ is given by  
\begin{eqnarray}
  && \tr{ \left( \CS_3^{(a)}P_+^{(b)} + \CA_3^{(a)}P_-^{(b)}
                   +P_+^{(a)}\CS_3^{(b)} + P_-^{(a)}\CA_3^{(b)} \right) \Delta }
               \nonumber \\
  &=& \frac{1}{2}\lambda_+
              \Big( \dim V_\CS^{(a)}\dim V_\CM^{(b)}+\dim V_\CA^{(a)}\dim V_\CM^{(b)} 
                                    \nonumber \\
  && \hspace{5ex} +\dim V_\CM^{(a)}\dim V_\CS^{(b)}+\dim V_\CM^{(a)}\dim V_\CA^{(b)} \Big).
            \label{not_mm_part}
\end{eqnarray}

When the total state belongs to $V_\CM^{(a)}\otimes V_\CM^{(b)}$, construction of the best 
strategy for Alice and Bob is rather involved.  
First we introduce the following operators $X_1$ and $X_2$ for each of Alice's space and Bob's space:  
\begin{eqnarray}
  X_1^{(\kappa)} &=& \frac{2}{\sqrt{3}} D^{(\kappa)},  \nonumber \\
  X_2^{(\kappa)}&=& 2A^{(\kappa)},\ \ \ (\kappa=a,b).
\end{eqnarray} 
Note that $X_1^{(\kappa)}$ and $X_2^{(\kappa)}$ anticommute and 
$(X_1^{(\kappa)})^2=(X_2^{(\kappa)})^2=1$ in the mixed symmetric space $V_\CM^{(\kappa)}$.  
The operator $\Delta$ in terms of $X_i^{(\kappa)}$ is not diagonal with respect to 
the index $i$. We further define rotated $X_i$'s in order to diagonalize $\Delta$ with 
respect to the index $i$. 
\begin{eqnarray}
  Y_1^{(\kappa)} &=& \cos\theta X_1^{(\kappa)} + \sin\theta X_2^{(\kappa)}, 
                          \nonumber \\
  Y_2^{(\kappa)} &=& -\sin\theta X_1^{(\kappa)} + \cos\theta X_2^{(\kappa)},\ \ (\kappa=a,b).
\end{eqnarray}
We find that $\Delta$ takes the following "diagonal" form: 
\begin{eqnarray}
  \Delta = \frac{1}{2} \left( \eta_1-\eta_2 + 
              \lambda_+Y_1^{(a)}Y_1^{(b)} + \lambda_-Y_2^{(a)}Y_2^{(b)} \right),
\end{eqnarray}
if we take
\begin{eqnarray}
  \cos 2\theta &=& \frac{\eta_1-\eta_2}{2\sqrt{1-\eta_1\eta_2}}, \\
  \sin 2\theta &=& \frac{\sqrt{3}}{2\sqrt{1-\eta_1\eta_2}}. 
\end{eqnarray}
Eigenvalues of $Y_i^{(\kappa)}$ are 1 and -1 with multiplicity $\dim V_\CM^{(\kappa)}/2$ 
since we have  
\begin{eqnarray}
  && (Y_1^{(\kappa)})^2 = 1,\ (Y_2^{(\kappa)})^2 = 1, \nonumber \\
  &&  Y_1^{(\kappa)}Y_2^{(\kappa)}+Y_2^{(\kappa)}Y_1^{(\kappa)} = 0.
\end{eqnarray}
And the positive- and negative-eigenvalue subspaces of $Y_1^{(\kappa)}$ are transformed to 
each other by the operation of $Y_2^{(\kappa)}$ and vice versa. We should also notice that 
$|\lambda_-| \ge |\lambda_+|$ when $\eta_1 \le \eta_2$. 
These considerations imply that the optimal separate POVM element is given by 
$Q_+^{(a)}\otimes Q_-^{(b)} + Q_-^{(a)}\otimes Q_+^{(b)}$, where 
$Q_\pm^{(\kappa)}$ is the projector onto the positive- and negative-eigenvalue 
subspace of $Y_2^{(\kappa)}$. The contribution to $\tr{E_1^{{\rm L}}\Delta}$ 
is found to be 
\begin{eqnarray}
  & & \tr{\left(Q_+^{(a)}\otimes Q_-^{(b)} + Q_-^{(a)}\otimes Q_+^{(b)}\right)\Delta}
                          \nonumber \\
  &=& \frac{1}{4}(\eta_1-\eta_2-\lambda_-) \dim V_\CM^{(a)} \dim V_\CM^{(b)} 
                          \nonumber \\
  &=& \frac{1}{4} \lambda_+ \dim V_\CM^{(a)} \dim V_\CM^{(b)},
               \label{mm_part}
\end{eqnarray}
where we used $\tr{Q_\pm^{(\kappa)}Y_1^{(\kappa)}}=0$. 

Thus the whole POVM element is given by 
\begin{eqnarray}
  E_1^{{\rm L}} &=& \CS_3^{(a)}P_+^{(b)} + \CA_3^{(a)}P_-^{(b)}
                   +P_+^{(a)}\CS_3^{(b)} + P_-^{(a)}\CA_3^{(b)} 
                                 \nonumber \\
                & &+Q_+^{(a)}Q_-^{(b)} + Q_-^{(a)}Q_+^{(b)}.   \label{E1L}
\end{eqnarray}
Adding Eq.(\ref{not_mm_part}) and Eq.(\ref{mm_part}), we find that 
$\tr{E_1^{{\rm L}}\Delta}$ indeed attains the maximum value given by 
the global POVM element $E_1=P_+$:
\begin{eqnarray}
   \tr{E_1^{{\rm L}}\Delta} = \frac{1}{2}\lambda_+\dim{V_\CM} = \tr{P_+\Delta}.
\end{eqnarray} 
To show the above equality, we used the relation
\begin{eqnarray}
  \dim{V_\CM} &=& \dim{V_\CS^{(a)}}\dim{V_\CM^{(b)}}+\dim{V_\CM^{(a)}}\dim{V_\CS^{(b)}}
                                   \nonumber \\
              & &+\dim{V_\CA^{(a)}}\dim{V_\CM^{(b)}}+\dim{V_\CM^{(a)}}\dim{V_\CA^{(b)}}
                                   \nonumber \\
              & &+\frac{1}{2}\dim{V_\CM^{(a)}}\dim{V_\CM^{(b)}},
\end{eqnarray}
which can be readily verified by a straightforward calculation. 
The factor $1/2$ in front of $\dim{V_\CM^{(a)}}\dim{V_\CM^{(b)}}$ reflects the 
fact that the inner product (Kronecker product) of two mixed symmetric representations 
contains the totally symmetric and antisymmetric representations in addition to 
the mixed symmetric representation.

On the other hand, we can show that the POVM element $E_1^{{\rm L}}$ given in Eq.(\ref{E1L}) 
can be implemented with an LOCC protocol. 
First Alice and Bob determine which permutation symmetries each one's local state has; 
totally symmetric, totally antisymmetric, or mixed symmetric. 
If one of them finds that his or her state is totally symmetric or antisymmetric and 
the other party's state is mixed symmetric, this party with the mixed symmetric state 
performs the measurement by the projectors $\{P_+^{(\kappa)},P_-^{(\kappa)}\}$. 
They conclude that the input state is $\rho_1$ if the combination of their 
outcomes is either "(symmetric,$P_+$)" or "(antisymmetric,$P_-$)". Otherwise 
they conclude that the input state is $\rho_2$. When Alice and Bob find 
both the local states are mixed symmetric, they perform the projection 
measurement by $\{ Q_+^{(\kappa)},Q_-^{(\kappa)}\}$. 
They conclude the input state is $\rho_1$, only when the combination of outcomes is 
"($+$,$-$)". 

Thus we conclude that the pure-state identification with an arbitrary {\it a priori} 
occurrence probability can be optimally performed within LOCC scheme.

\section{Concluding remarks}
It has been known that two bipartite pure states can be optimally discriminated 
within LOCC scheme if classical knowledge on the two states are available. 
In this paper, we showed that this is also true in the identification problem 
of two bipartite pure states, where no classical knowledge on the two states 
is given but only a copy of the two states is available as reference states.  

We assumed the number $N$ of copies of each state is one.  
In the limit of large $N$, the identification problem reduces to the 
standard discrimination problem. This is because one can obtain complete 
classical information on the reference states by performing a tomographical 
measurement on infinitely many copies of them. Therefore, it has been shown 
that the pure-state identification can be optimally performed by means of LOCC 
when $N=1$ and $N=\infty$. We conjecture that this is also true for arbitrary $N$. 

In this paper we allowed Alice and Bob to make a mistake in identifying 
the input state with one of the reference states. Instead we can consider 
a different version of identification problem, unambiguous (conclusive) 
identification problem \cite{Bergou05,Hayashi06}, where one is not allowed to 
make a mistake. It is of interest to ask whether the unambiguous identification 
can be performed optimally by means of LOCC and the results on this issue will 
be discussed elsewhere \cite{Hayashi08}.


\begin{thebibliography}{}
\bibitem{Helstrom76}
C.~W.~Helstrom,
{\it Quantum Detection and Estimation Theory} 
(Academic Press, New York, 1976).

\bibitem{Ivanovic87}
I.~D.~Ivanovic,
Phys. Lett. A {\bf 123}, 257 (1987).

\bibitem{Dieks88}
D.~Dieks,
Phys. Lett. A {\bf 126}, 303 (1988).

\bibitem{Peres88}
A.~Peres,
Phys. Lett. A {\bf 128}, 19 (1988).

\bibitem{Wootters82}
W.~K.~Wootters and W.~H.~Zurek,
Nature {\bf 299}, 802 (1982). 

\bibitem{Bennett99}
C.~H.~Bennett, D.~P.~DiVincenzo, C.~A.~Fuchs, T.~Mor, 
E.~Rains, P.~W.~Shor, J.~A.~Smolin, and W.~K.~Wootters, 
Phys. Rev. A{\bf 59}, 1070 (1999).

\bibitem{Koashi07}
Masato Koashi, Fumitaka Takenaga, Takashi Yamamoto, and Nobuyuki Imoto, 
arXiv:0709.3196v1 [quant-ph]. 

\bibitem{Walgate00}
Jonathan~Walgate, Anthony~J.~Short, Lucien~Hardy, and Vlatko~Vedral,
Phys. Rev. Lett. {\bf 85}, 4972 (2000).

\bibitem{Virmani01}
S.~Virmani, M.~F.~Sacchi, M.~B.~Plenio, and D.~Markham,
Phys. Lett. A{\bf 288}, 62 (2001).

\bibitem{Chen01}
Y.-X.~Chen and D.~Yang, 
Phys. Rev. A{\bf 64}, 064303 (2001)

\bibitem{Chen02}
Y.-X.~Chen and D.~Yang, 
Phys. Rev. A{\bf 65}, 022320 (2002)

\bibitem{Ji05}
Zhengfeng~Ji, Hongen~Cao, and Mingsheng~Ying,
Phys. Rev. A{\bf 71}, 032323 (2005). 

\bibitem{Hayashi05}
A.~Hayashi, M.~Horibe, and T.~Hashimoto, 
Phys. Rev. A{\bf 72}, 052306 (2005). 

\bibitem{Bergou05}
Janos~A.~Bergou and Mark~Hillery,
Phys. Rev. Lett. {\bf 94}, 160501 (2005).

\bibitem{Hayashi06}
A.~Hayashi, M.~Horibe, and T.~Hashimoto,  
Phys. Rev. A{\bf 73}, 012328 (2006). 

\bibitem{Bergou06}
Janos~A.~Bergou, Vladimir~Buzek, Edgar~Feldman, Ulrike~Herzog, and Mark~Hillery,
Phys. Rev. A{\bf 73}, 062334 (2006).

\bibitem{Hayashi04}
A.~Hayashi, T.~Hashimoto, and M.~Horibe, 
Phys. Rev. A{\bf 72}, 032325 (2005). 

\bibitem{Hamermesh62}
M.~Hamermesh,
{\it Group Theory and its Application to Physical Problems},
(Addison-Wesley, Reading, MA, 1962).

\bibitem{Hayashi08}
A.~Hayashi, Y.~Ishida, T.~Hashimoto, and M.~Horibe, 
arXiv:0801.0128v1 [quant-ph].

\end{thebibliography}
\end{document}